\documentclass[aps,onecolumn,superscriptaddress]{revtex4} 

\usepackage{graphicx,amssymb,amsmath}
\usepackage{color}
\usepackage{rotating}
\usepackage{ulem}

\newcommand{\be}{\begin{equation}}
\newcommand{\ee}{\end{equation}}

\newcommand{\bea}{\begin{eqnarray}}
\newcommand{\eea}{\end{eqnarray}}

\newcommand{\br}{\mathbf{r}}
\newcommand{\bR}{\mathbf{R}}
\newcommand{\tbr}{{\tilde{\bf r}}}

\newcommand{\bxi}{\mbox{\boldmath${\xi}$}}
\newcommand{\bt}{\mathbf{t}}

\newcommand{\tU}{\tilde U}
\newcommand{\tD}{\tilde D}
\newcommand{\tR}{\tilde R}
\newcommand{\tk}{\tilde \kappa}

\def\rpar{\mathbf{r}_\parallel}

\def\eq#1{Eq.~(\ref{#1})}

\begin{document}

\title{Statistical physics and mesoscopic modeling to interpret tethered particle motion experiments}
\author{Manoel Manghi}
\email{manghi@irsamc.ups-tlse.fr}
\affiliation{Laboratoire de Physique Th\'eorique (IRSAMC), Universit\'e de Toulouse, CNRS, UPS, France}
\author{Nicolas Destainville}
\email{destain@irsamc.ups-tlse.fr}
\affiliation{Laboratoire de Physique Th\'eorique (IRSAMC), Universit\'e de Toulouse, CNRS, UPS, France}
\author{Anna\"el Brunet}
\email{annaelle.brunet@medisin.uio.no}
\affiliation{Department of Molecular Medicine, Institute of Basic Medical Sciences, Faculty of Medicine, University of Oslo, 0317 Oslo, Norway}
\date{\today}

\begin{abstract}
Tethered particle motion experiments are versatile single-molecule techniques enabling one to address \textit{in vitro} the molecular properties of DNA and its interactions with various partners involved in genetic regulations. These techniques provide raw data such as the tracked particle amplitude of movement, from which relevant information about DNA conformations or states must be recovered. Solving this inverse problem appeals to specific theoretical tools that have been designed in the two last decades, together with the data pre-processing procedures that ought to be implemented to avoid biases inherent to these experimental techniques. These statistical tools and models are reviewed in this paper.  
\end{abstract}

\maketitle

\section{Introduction}

The main advantage of single-molecule techniques over traditional bulk experiments is the possibility to disentangle sample heterogeneity and to gain insight into subpopulation properties. The tethered particle motion (TPM\footnote{Principal abbreviations used in this work: TPM:  tethered particle motion; htTPM: high-throughput TPM; MTT: magnetic torque tweezers; AFM: atomic force microscopy; HMM: hidden Markov model.}) 
single-molecule technique has been developed in the early 1990's~\cite{Schafer1991,Yin1994} to detect and quantify  conformational changes of biopolymers induced by their interaction with other molecular partners or changes in their environment~\cite{Finzi1995,Zocchi2006,Fan2018}. It consists in tracking the Brownian motion of a nano-particle (tens to hundreds of nanometers in diameter) attached to a glass surface by a biopolymer such as a DNA molecule and measuring the particle amplitude of movement and its changes when experimental conditions are modified. TPM experiments do not require expensive experimental set-ups, in particular because the particle is tracked by an optical microscope. This explains why numbers of experimental groups adopt this technique to investigate the effects of agents (e.g., enzymes, drugs, ions, or more generally $p$H, ionic strength or temperature) acting on the characteristics of the tethering polymer, such as its persistence length, its conformation, or its denaturation properties. A variant of TPM is Tethered fluorophore motion (TFM)~\cite{Pinkney2012,May2014}. It uses the same principles as TPM but employs a fluorophore (and sometimes two~\cite{Schickinger2018}) in place of the particle. TFM can thus be combined with fluorescence techniques such as F\"orster resonance energy transfer. However, TFM is limited in observation time because of fluorophore photobleaching~\cite{May2014}. 

Optical and magnetic tweezers~\cite{Bustamante2003,Neuman2008} are another class of powerful tools to investigate the elastic properties of DNA molecules. Optical tweezers~\cite{Smith1996,Wang1997} rely on a focused laser beam to provide an attractive force on the order of the pico-Newton (pN) to manipulate micrometric particles. Magnetic tweezers~\cite{Smith1992,Gosse2002} consist of two permanent magnets  producing a horizontal magnetic field at the location of a magnetic particle. In both cases, as in TPM, the particle is attached to a biopolymer tether, itself grafted to the glass surface. Magnetic torque tweezers (MTT) are an extension of conventional magnetic tweezers where a cylindrical magnet creates a vertical magnetic field and permits to apply both forces and torques. At zero turn the particle is at its rotational equilibrium position and the tethered DNA is torsionally relaxed. After applying turns the DNA molecule is twisted, which gives access to the torsional elastic properties of DNA and also its non-linear response when twist is converted into writhe through the creation of superhelical DNA regions~\cite{Marko1994}. 

As compared to optical or magnetic tweezers, no external force is applied to the particle in TPM~\cite{Zocchi2006}, if not the weak repulsion exerted by the glass surface on the polymer and the bead, in the tens of fN range~\cite{Segall2006}. Studying the biopolymer in quasi-force-free conditions enables one to tackle its equilibrium properties, as well as reaction rates between two (or more) states such a assembly/disassembly rates of DNA constructs or binding/unbinding rates of enzymes on a tensionless DNA~\cite{Manghi2010}. These different techniques are thus complementary. 

Recent improvements of TPM rely on multiplexing, hundreds of DNA-bead complexes being positioned in a controlled manner by soft nano-lithography and monitored in parallel. The ensuing technique is called high-throughput TPM (htTPM)~\cite{Plenat2012}. By drastically reducing acquisition time as compared to anterior setups where the molecules where observed one by one, multiplexing gives access to highly refined statistics allowing one to distinguish between closely related conformations~\cite{Brunet2015,Brunet2015B,Fournes2016,Brunet2018,CT2019}. Dealing with these refined statistics justifies the development of improved statistical tools and modeling to interpret them, as detailed in this Review. 

From a biological perspective, single-molecule techniques have enabled many research groups to decipher key mechanisms at play in cells, starting with the pioneering paper by Schafer and coworkers in 1991, where the progressive shortening of the tether gave access to the processivity of immobilized RNA polymerases~\cite{Schafer1991}. The estimated value, even though rather rough at this time, was of a dozen of bps/s (see also Ref.~\cite{Yin1994}), in satisfying agreement with measurements in solution. Later, using a similar strategy, the RuvAB-directed branch migration of individual Holliday junctions~\cite{Dennis2004} was measured, also on the order of 10 to 20~bps/s, as well as its dependence on the construct sequence. Several other examples of biological applications can be found in previous review articles~\cite{Fan2018,Tardin2017} or will be discussed in the present Review.  The interpretation of these experiments not only depends on accurate measurements but also on adequate and reliable physical models able to account for sometimes very weak and subtle effects. It allows one to identify the relative roles of the intrinsic parameters of the system. The bead size, the tether length, the surface state of the substrate and the solvent are all likely to play pivotal roles in this context, as well as the acquisition rate of the camera used to track the particle. Their different contributions must be precisely quantified, as discussed below. This Review article can be seen as a companion article to the one by Jean-Fran\c cois Allemand, Catherine Tardin and Laurence Salom\'e in the same issue~\cite{CT2019}. It gives additional details about physical, mathematical and algorithmic issues related to TPM and the ways to tackle them. {Even though they are not the main focus of this Review, optical and magnetic tweezers or AFM often come as complementary tools to study single-DNA molecules under force and/or torque. They also need theoretical and algorithmic tools for the interpretation and modeling of experiments. The last decade has witnessed their rapid development, and the reader can for example refer to Refs.~\cite{Vilfan2009,Manghi2012,Heller2014,Truex2015,Dulin2015,Sarkar2016,Lin2017,Kriegel2017A} for further detail. When the connection with TPM experiments is meaningful, we shall discuss some works dealing with force and torque experiments in the present Review}.

\section{Modeling single DNA-molecule experiments and their dynamics}
\label{model}

Several coarse-grained models have been developed in the past decades to model a single DNA molecule. These models are either numerical and/or analytical, the simplest one being the Gaussian chain, for which the end-to-end distance is $R_{\rm ee}\equiv\sqrt{\langle \bR^2\rangle}\simeq (b L)^{1/2}$, a valid expression\footnote{The average $\langle\ldots\rangle$ is an ensemble average over realizations. When dealing with experiments, it will become an average over time, assuming the validity of the ergodic theorem.} as soon as the DNA contour length $L$ is much larger than the Kuhn length  $b=2\ell_p$. The DNA persistence length is $\ell_p=\kappa/(k_BT)$ where $\kappa$ is the bending modulus and is approximately $50$~nm for double-stranded DNA {in physiological conditions~\cite{Brunet2015}}. The associated numerical model is the bead-spring model which is easy to implement. It consists in modeling the DNA by $N$ beads, whose diameter is equal to the Kuhn length, connected by springs, in Brownian dynamics or Monte Carlo simulations. Although this model is central to understand polymer properties at large scales~\cite{deGennes1979}, it is not adapted to single-DNA molecule experiments, which are interested in DNA lengths from few hundreds to few thousand base-pairs, on the order of the Kuhn length. Moreover the bead has a radius much larger than the dsDNA radius $\simeq 1$~nm. It should be noted that at large scales, DNA is in good solvent conditions, i.e. the DNA is swollen compared to the Gaussian chain. The end-to-end distance is now $R_{\rm ee} \simeq (v/b)^{1/5}L^{3/5}$ (for $L\gg b$) where the excluded volume $v$ depends on the salt concentration.

The most adapted model is the worm-like (or semi-flexible) chain model because it reveals  the mechanical and statistical properties that are probed in single-molecule experiments, without describing the DNA structure in detail. This model considers the chain as a homogeneous stiff rod, the bending energy of which leads to a short-ranged tangent-tangent correlation, $\langle \bt(s)\cdot\bt(0)\rangle = e^{-s/\ell_p}$ where $s$ is the curvilinear index along the chain and $\bt(s)$ the normalised tangent vector. The end-to-end distance is then given by the Kratky-Porod result~\cite{Doi1986}
\be
R_{\rm ee}^2\simeq 2\ell_p^2\left(\frac{L}{\ell_p}-1+e^{-L/\ell_p}\right)
\label{KP}
\ee
which yields the two good limits of the rigid rod, {$R_{\rm ee}^2\simeq L^2$,} when $L\ll\ell_p$ and the Gaussian chain, {$R_{\rm ee}^2\simeq 2 L \ell_p$,}  when $L\gg\ell_p$.
The associated discretised numerical model is a bead-spring model with a large spring stiffness to enforce the chain connectivity and an additional bending energy $E_b=\kappa\sum_{i=1}^{N-1} (1-\cos\theta_i)$ where $\kappa=\ell_p k_B T$ is the bending modulus and $\theta_i$ the angle between two consecutive links.
In this discrete worm-like chain model, the successive beads are free to rotate with respect to each other (which is equivalent to set the torsional modulus $C$ to 0). {Note that in the torque experiments, torsion must also be taken into account: the twist angle  $\phi_i$ is defined between two consecutive base-pairs (or by defining a material frame for each bead) and the associated elastic energy of a torsional spring  writes, in the limit of large $C$ (valid for DNA), $E_t=\frac{C}2\sum_{i=1}^{N-1} (\phi_i-\phi_0)^2$, where $\phi_0=0.62$~rad is the equilibrium DNA twist  in physiological conditions).}

Depending on the experimental setup, possible boundary conditions and/or interactions between the DNA and external objects might be considered. In TPM, one DNA end is tethered to a substrate, which is usually modeled as a freely rotative joint. The other DNA end is attached to a spherical particle. The DNA-particle link is also treated as freely rotative joint, except in torsion experiments. The glass coverslip is treated as a hard wall boundary condition limiting the motion of the DNA and the particle to the upper half plane. Although easy to implement numerically, the hard wall condition and the large particle size modify the equilibrium statistics of the DNA in TPM experiments. The amplitude of movement of the particle is defined as $\sigma=\sqrt{\langle\br_{\parallel}^2\rangle}$ where $\br_{\parallel}$ is the two-dimensional particle position parallel to the coverslip. Note that without loss of generality, we have set here $\langle \rpar(s) \rangle=0$. The amplitude of movement $\sigma$ can be estimated analytically only in the limit of flexible DNA ($L\gg\ell_p$)~\cite{Segall2006,Manghi2010,Lindner2011}, and some interpolation formulas have been proposed in the semi-flexible regime~\cite{Nelson2006}.

Concerning the dynamical properties of the DNA, the relaxation time (see next Section) is also modified by the experimental setup. In particular, the no-slip boundary condition enforced by the presence of the coverslip slows down the DNA--particle dynamics. Hence in the numerical simulations, the hydrodynamics interactions induced by the wall are encoded using Fax\'en's law {prescribing how the diffusion coefficient of both the DNA molecule and the particle is reduced close to the wall, in order to satisfy the no-slip condition for the solvent velocity field at the wall}~\cite{Manghi2010}.

\begin{figure}[ht]
\begin{center}
\includegraphics[height=6.1cm]{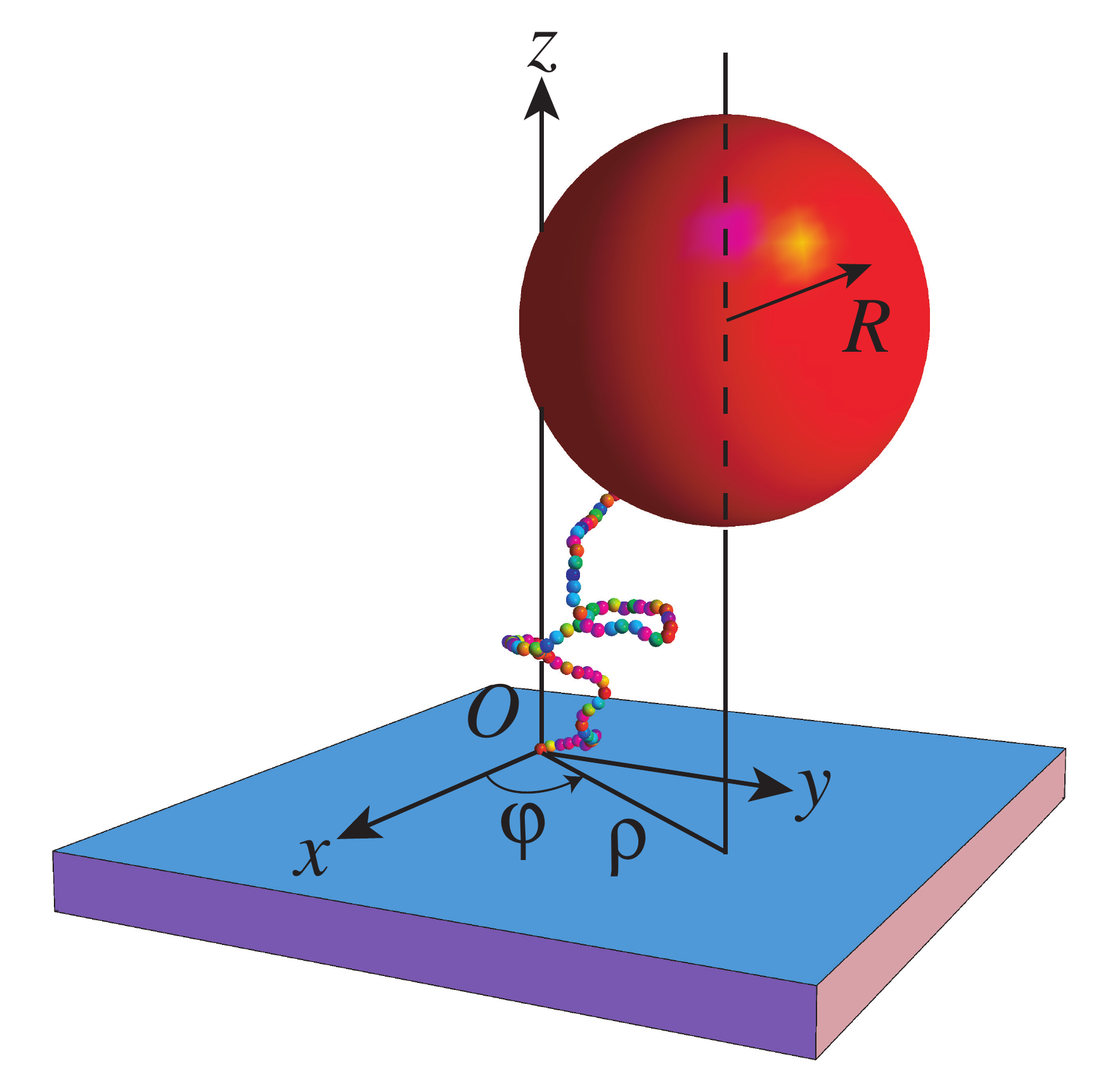} 
\caption{A TPM numerical model: the DNA molecule is modeled as a polymer chain made of $N$ connected beads (various colours), anchored to the coverglass (in blue) at one extremity and to the tracked particle (in red) at the other end. The 2D position $\rpar$ of the particle center is represented by the polar coordinates $\rho$ and $\varphi$ in the $(xOy)$ plane.
\label{TPMsketch}}
\end{center}
\end{figure}

\section{Dealing with statistical and systematic errors}
\label{data:processing}

We first introduce the different experimental times of interest, here and in the sequel. The camera acquisition period is denoted by $T_{\rm ac}$. It will play an important role when dealing with the blurring effect below. It typically ranges from few ms for fast acquisition devices~\cite{Kumar2014} to few tens of ms at video rate. The camera exposure time is $T_{\rm ex}\leq T_{\rm ac}$, {whilst in general $T_{\rm ex} = T_{\rm ac}$.} It must be shorter than the characteristic physical times of interest in the experiment {in order to have access to all relevant events}. As for the trajectory duration, it must be long as possible in order to reduce statistical uncertainties. Since the tethered particle is not subject to photobleaching, the trajectory can be recorded for several minutes in TPM, which is a great advantage in terms signal-to-noise ratio as compared to single-fluorophore techniques.  

One of the objectives here is to measure the useful correlation (or relaxation) time $\tau$ of the 2D tethered-particle position $\mathbf{r}_\parallel$. It sets the typical time needed for the particle to explore its configuration space. It is defined through the auto-correlation function
\begin{equation}
C(t) = \langle \rpar(s+t)\cdot \rpar(s) \rangle - \langle \rpar(s) \rangle^2
\label{Ct}
\end{equation}
where the average {$\langle \ldots \rangle$} is taken along the trajectory. For sake of simplicity, we assume that the time correlation function has the form $C(t)=C(0) e^{-t/\tau_{\rm m}}$. This expression is exact only in the case of a quadratic confining potential. {In the case of a more complicated confining potential induced by the polymer tether}, it is only an approximation because the auto-correlation function is a sum of decreasing exponentials and $\tau_{\rm m}$ is then associated with the slowest diffusion mode, which dominates at long times~\cite{Doi1986}. In practice, we shall see below that the measured correlation time $\tau_{\rm m}$ is slightly larger than the real one $\tau$ because of blurring effets that we aim at quantifying in the present section. 

\subsection{Cleaning data from spurious points}
 
A first data preprocessing step is essential to minimize the contributions of unwanted biases and experimental variability.  First of all, it is essential to deal with the measurement heterogeneity induced by undesirable artefacts due, e.g., to ill-assembled objects. For all single molecule approaches involving anchored DNA molecules, the substrate surface state is crucial~\cite{Zocchi2006}. This becomes even more critical when studying dynamically acting proteins or molecules binding on the DNA, and especially for AFM~\cite{Ando2013} or tweezer~\cite{Gietl2013} manipulations. Such approaches require that the DNA molecules binds to substrate surface without modifying physiological functions and properties, and that only desired anchoring is realized, without any sort of spurious secondary attachments~\cite{Manghi2010}.

When advanced experimental protocols are not sufficient to prevent those artefacts, statistical tools offer an alternative way to do so during the data preprocessing step. In TPM experiments, malformed DNA-particle complexes, e.g. with two grafted DNA molecules instead of an expected unique one would interfere with main population of interest and hamper the authentic experimental noise. Applying a selecting filter based on the asymmetry factor of the 2D trajectories allows one to select only well-defined tethered DNA/particle complexes as mentioned in the article by Allemand \textit{et al.} in this issue~\cite{CT2019}. {An asymmetry factor (or aspect ratio) larger than 1.35 is assumed to be associated with two DNA tethers cross-linked to the same tracked particle~\cite{Brunet2015}.}

{However, this criterion may appear to be insufficient to deal with spurious non-specific binding of some particles to the coverslip. One must get rid of trajectories laying in the far tails of the amplitude of movement distribution. {Hence}, if a single population is expected, for example when extracting a bending angle or the persistence length (see Section~\ref{ex:sec} below), trajectories with amplitude of movement outside the interval (mean $\pm 2.5$ standard deviations) were discarded in Refs~\cite{Brunet2015,Brunet2015B,Brunet2018}.}

Extracting the particle displacement-time properties provide an {alternative and complementary} objective criterion for classifying the trajectories into the main population and \textit{outliers}. The correlation time for each 2D TPM-trajectories, $\tau$, is determined as described above. The resulting data for the global DNA population is expected to behave as a single population for a given DNA state. Based on that, {and as above for amplitude of movements, a trajectory} is declared as an outlier if its $\tau$ lays in the far tails of this distribution and is then sorted out. In practice, sorting out points deviating by more than 1 or 2 standard deviations from the average value is {also}  a reasonable criterion~\cite{Brunet2015B}. Based on the same principle for filtering experimental data, Schickinger \textit{et al.}~\cite{Schickinger2018} exploit the average dwell-time. It is determined in each specific state, unbound and bound DNA, for each particle. Comparing bound vs. unbound dwell times reveals multiple data point clusters and provided a criterion of selection. Considerable discrepancy with the main populations leads to the discrimination of outliers.

More generally, all single-molecule approaches come with their own specific protocol-induced defects, which need to be taken into account in order to clean data from outliers before confronting them to statistical analysis and theoretical modeling. This is even more true when using high-throughput approaches that provide high sampling levels and allow to define precisely the main conformation populations.

\subsection{Subtracting instrumental drift}

Once data have been cleaned from outliers, the first systematic error to correct comes from the instrumental drift, due, \textit{inter alia}, to thermal expansion of the observation setup~\cite{Manghi2010}. Along a given trajectory, the DNA approximate anchoring point at each time $t_0$ is determined by averaging the particle position over an interval of duration $T_{\rm av}$ (typically 1 or 2~s, sometimes even more~\cite{Johnson2014}) centered at $t_0$, and then subtracted from $\mathbf{r}_\parallel$. If $T_{\rm av}$ is chosen to be much larger than the measured relaxation time $\tau_{\rm m}$, this anchoring point is determined with a good accuracy. This sets $\langle \mathbf{r}_\parallel \rangle \simeq 0$, as desired.

Note that subtraction of drift induces {non-trivial} systematic anti-correlations at short times $t\ll T_{\rm av}$. Indeed, let us assume that the measured time correlation function in absence of drift also has the form $C_{\rm m}(t)=C_{\rm m}(0) e^{-t/\tau_{\rm m}}$. Then subtraction of drift modifies it to 
\begin{equation}
\frac{C_{\rm m}(t)}{C_{\rm m}(0)} = \left(1+2\frac{\tau_{\rm m}}{T_{\rm av}} \right)e^{-t/\tau_{\rm m}} - 2\frac{\tau_{\rm m}}{T_{\rm av}}
\label{drift:corr}
\end{equation}
at first order in $2\tau_{\rm m}/T_{\rm av} \ll 1$~\cite{Manghi2010}. This will have to be taken into account when measuring  $\tau_{\rm m}$ below. 

\subsection{Correcting blurring effect}
\label{blur}

Another important source of systematic error in TPM comes from the finiteness of the camera exposure time $T_{\rm ex}$. An image in fact represents the optical signal averaged over a time interval of duration $T_{\rm ex}$. In Refs.~\cite{Destainville2006,Nelson2006,Towles2009,Manghi2010}, the ensuing time-averaging (or blurring) effect in single-particle tracking experiments was investigated. It occurs whenever the trajectory of the tracked particle is confined in a bounded domain, not only in TPM but also when tracking plasma membrane constituants, for instance. Indeed, in the extreme case where the frame exposure time would be much larger than the system auto-correlation time, itself inversely proportional to the domain area (see below), the measured particle position during this period would remain very close to the anchoring point, giving the erroneous impression that the amplitude of motion is much smaller than its actual value. However, we shall see that when it is not too strong, this effect can efficiently be corrected. 

By using Eq.~\eqref{drift:corr}, the measured correlation function $C_{\rm m}(t)$ is first fitted to obtain the measured correlation time $\tau_{\rm m}$, the only free parameter in this expression (the value of $T_{\rm av}$ has been chosen from the beginning). It can be proven that the real correlation time $\tau$ is given by 
\begin{equation}
\tau \simeq \tau_{\rm m} - \frac{T_{\rm ex}}3
\label{corr:time}
\end{equation}
which remains a correct approximation while $\tau_{\rm m} \geq 2\;T_{\rm ex}/3$~\cite{Destainville2006,Manghi2010}. From this value, the  diffusion domain size can now be corrected. It is characterized by the trajectory standard deviation $\sigma \equiv \sqrt{\langle \rpar^2 \rangle}$, also called ``amplitude of movement'', measured on a sufficiently long interval in order  to accurately sample configurations~\cite{Kumar2014} (see Section~\ref{threshold}). If $\sigma_{\rm m}$ is its measured value, then the real one is recovered in its turn from
\begin{equation}
\sigma \simeq \sigma_{\mathrm{m}} \left[2 \frac{\tau}{T_{\rm ex}} - 2 \left(\frac{\tau}{T_{\rm ex}}\right)^2 \left(1 - e^{-\frac{T_{\rm ex}}{\tau}} \right)\right]^{-1/2}.\label{ell:ellm0}
\end{equation}
For example, if $T_{\rm ex} =2 \tau$ then $\sigma_{\mathrm{m}} \simeq  0.75 \sigma$~\cite{Destainville2006}. In the case where $T_{\rm ex} \ll \tau$, we naturally get $\sigma \simeq  \sigma_{\mathrm{m}}$. Note that the particle diffusion coefficient $D$ depends on both $\sigma$ and $\tau$, through $D={\rm Const.}\, \sigma^2/\tau$, where ${\rm Const.}$ depends on the domain geometry~\cite{Bickel2007}. The measured value $D_{\rm m}$ can be substantially different from the actual one $D$, e.g., $D_{\rm m} \simeq 0.34 D$ if  $T_{\rm ex} =2 \tau$. It is necessary to correct both $\sigma$ and $\tau$ thanks to the above formulae to get the correct value of $D$. 

Incorrectly dealing with this blurring phenomenon can have dramatic effects when varying the experimental temperature $T$. Indeed, the water viscosity $\eta_{\rm w}(T)$ decreases rapidly with increasing $T$. This leads to a 4-fold fall of $\tau \propto D^{-1} \propto \eta_{\rm w}(T)/T$ when $T$ grows from 15 to 70$^\circ$C~\cite{Brunet2018}. Even though larger than $T_{\rm ex}$ at low $T$, $\tau$ likely becomes comparable to or smaller than $\tau$ at high $T$, then requiring correction of the blurring phenomenon. This issue has previously led to erroneous conclusions about denaturation profiles of DNA as observed by TPM~\cite{Driessen2012,Brunet2018} (see also  Section~\ref{TvsLp} below). 

\subsection{Estimating error bars}

The last step is to determine average values and associated error bars. The most straightforward way is to estimate the single average of an observable $\mathcal{O}$ over the distribution of the set of $N$ data point as its mean $\mu_\mathcal{O} \equiv \langle \mathcal{O} \rangle$. Statistical fluctuations are estimated through the the variance $\sigma_\mathcal{O}^2 \equiv \langle \mathcal{O}^2 \rangle - \langle \mathcal{O} \rangle^2$, and the standard deviation $\sigma_\mathcal{O}$. Assuming statistical independence of samples, the error bar on $\mu_\mathcal{O}$ is then $\sigma_\mathcal{O} / \sqrt{N}$ (68\% confidence interval) or twice this value (95\% confidence). 

More advanced methods can be used as the \textit{jackknife}, used in Refs.~\cite{Imparato2008,Loong2012,Tapia2017}, or the \textit{bootstrap}, used in Refs.~\cite{Konig2013,Brunet2015,Brunet2015B,Fournes2016,Brunet2018}, to estimate the error bar. Both are resampling methods. In the jackknife, one considers $N$ resampled sets of data, each containing all but one of the original data points. The bootstrap uses $M$ sets of data, each containing $N$ data points obtained by random sampling, performed by Monte Carlo, of the original set of $N$ points. During the Monte Carlo sampling with replacement, the probability that a data point is selected is $1/N$. The number of bootstrap samplings, $M$, should be chosen to be large enough so that the average bootstrap sampling is reproducible with sufficient accuracy. The jackknife approach leads to identical results each time it is run on the same set of data, which is not true for bootstrap.

\section{Solving the inverse problem}
\label{inverse:pb}

``Solving the inverse problem'' can be {generically} stated as {calculating, from a set of experimental measurements, the causes that produced them}. The aim is to gain insight into the physical properties of a system by indirect measurements or conversely to set up a predictive model that can reproduce observations. In the present context, coupling theory and experiments offers appropriate tools to probe the intrinsic physical properties of the DNA macromolecule: for instance, DNA persistence length or its local defects, from the apparent end-to-end distance of the polymer as accessible from TPM.

\subsection{Analytical approaches and their limits}
\label{extension}

The exact equilibrium distribution, $p(\br_{\parallel})$, of the amplitude of movement of the TPM particle can only be computed in the rigid and flexible limits~\cite{Manghi2010,Segall2006}. Using the mirror reflection argument, the probability distribution in the Gaussian (flexible) regime and in the limit where the particle radius is large, $R\gg \sqrt{L\ell_p}$, is given by~\cite{Manghi2010}:
\be
p_{\rm G}(|\br_{\parallel}|) \approx \sqrt{\frac{3}{\pi L\ell_p}} \frac{|\br_{\parallel}|}{\sqrt{\br_{\parallel}^2+R^2}} \exp\left[-\frac{3}{4L\ell_p} \left(\br_{\parallel}^2+2R^2-2R\sqrt{\br_{\parallel}^2+R^2}\right)\right].
\label{pGausslargeR}
 \ee
 The intermediate semi-flexible regime, $L\simeq\ell_p$, of interest in TPM  experiments, is well described by the worm-like chain model. For instance, in the experiments described in Ref.~\cite{Manghi2010}, DNA lengths vary between 400 and 2080~bp, which corresponds to $2<L/\ell_{\rm p}<14$. This model can be tackled analytically~\cite{Samuel2002,Stepanow2002} but the boundary conditions must be handled with care and the final step should be solved numerically, which does not provide any analytical formula for the probability distribution with fitting parameters. Moreover, it should be kept in mind that real chains are self-avoiding and that the presence of the {labelling} particle renders the problem even more intricate. 
For instance, the effect of the excluded volume of the particle is to widen and shift to large $|\br_{\parallel}|$ the Gaussian distributions. It then becomes analytically intractable for finite chains, but it can be tackled numerically by Monte Carlo or Brownian dynamics simulations (see Section~\ref{MC:MD}).

The {over-damped} dynamics of the DNA in the flexible regime is controlled by the Rouse time $\tau_{\parallel}=N R_G^2/D_0$ where $R_G$ is the radius of gyration and $D_0=k_BT/(6\pi\eta \ell_p)$  is the diffusion coefficient of a monomer sphere of radius $\ell_p$ in a liquid of viscosity $\eta$~\cite{Qian2000,Manghi2010}. However, when the TPM particle is grafted at one end, the relaxation time and the diffusion coefficient measured by tracking the particle dynamics are features of the dynamics of the whole DNA--particle complex. Using the Langevin equation, it is shown that the diffusion coefficient $D_c$ is given by:
\be
D_c= \frac{D_{\rm part}D_{\rm DNA}}{D_{\rm part}+D_{\rm DNA}}
\ee
where $D_{\rm part}=k_BT/(6\pi\eta R)$. Thus the particle does not slow down the complex provided that $D_{\rm part}\gg D_{\rm DNA}$. Knowing $D_{\rm part}$, the value of $D_{\rm DNA}$ is inferred from the measurement of $D_c$. Note however that these analytical considerations do not allow us to compare with TPM experiments due to (i) the fact that the chain is in the semiflexible regime; and (ii) the neglect of hydrodynamics interactions {in this approach}. 

\subsection{Numerical simulations}
\label{MC:MD}

Several types of numerical simulations have been developed to circumvent the above limitations. 
To test both the dynamics and the equilibrium properties in the TPM setup geometry, the adequate numerical methods are Brownian dynamics simulations and kinetic Monte Carlo simulations. 

In Brownian dynamics simulations, the evolution of each sphere position $\mathbf{r}_i(t)$ is governed by an iterative Langevin equation (discrete time step $\delta t$ and discrete time variable $n=t/\delta t$)
\be
\tbr_i(n+1)=\tbr_i(n)- \tD_0 \, \nabla_{\tbr_i}\tU(n) + \sqrt{2 \tD_0} \, \tilde{\bxi}_i(n),
\label{BD}
\ee
where the rescaled random displacement has variance unity $\langle\tilde{\bxi}_i(n)\cdot\tilde{\bxi}_j(m)\rangle=3\,\delta_{ij}\delta_{nm}$. The rescaled bare diffusion coefficient $\tD_0 = D_0\delta t/a^2$ is the diffusion constant in an unbounded space in units of the particle radius $a$ and time step $\delta t$. For sufficient numerical accuracy the usual choice is  $\tD_0 = 10^{-3}$--$10^{-5}$. The dimensionless potential $\tU=U/(k_BT)$ is the sum of stretching and bending potentials described in Section~\ref{model}. The excluded volume interaction is modeled by a repulsive, truncated Lennard-Jones potential $\tU_{LJ}=\sum_{i<j} [(b/|\tbr_i-\tbr_j|)^{12} - 2(b/|\tbr_i-\tbr_j|)^6+1]$ valid for separation $|\tbr_i-\tbr_j|<2$ and $b=\tR+1$ for $j=N$ and $b=2$ otherwise. Since the polymer motion is limited to the upper half-plane $z>0$, we use the reflection boundary condition: if a sphere intersects the substrate, its height $z_i$ is replaced by its mirror image $z_i^{\rm refl}=2a-z_i$ ($z_N^{\rm refl}=2(R-a)-z_N$ for the particle).

In Monte Carlo simulations, at each step (MCStep) $\delta t$, a bead is chosen uniformly at random among the $N+1$ possible ones (monomer spheres and labelling particle). Then a random move $\delta \mathbf{r}$ is attempted for this bead, uniformly in a ball of center 0 and radius $R_b$. {One shows that} in this case  $\langle \delta \mathbf{r}^2 \rangle =3R_b^2/5$. This quantity must be equal to $6 D_0 \delta t$, where $D_0$ is the diffusion coefficient of the spherical bead, depending on its diameter, which sets $R_b=\sqrt{10D_0\delta t}$. Interactions between adjacent beads are treated \textit{via} the potential $U$, whereas interactions between non-adjacent beads are of hard core nature, like surface-bead interactions: whenever a move would lead to the penetration of a bead into an other one or the surface, it is rejected. The physical time is incremented of $\delta t$ following each Monte Carlo Sweep (MCS, equal to a sequence of $N+1$ MCSteps). A simulation snapshot is shown in Fig.~\ref{TPMsketch} at equilibrium. Note that the choice of excluded volume interaction in Monte Carlo simulations saves computational time as compared to the calculation of truncated Lennard-Jones potentials used in Brownian Dynamics.

\begin{figure}[t]
\begin{center}
\includegraphics[height=6cm]{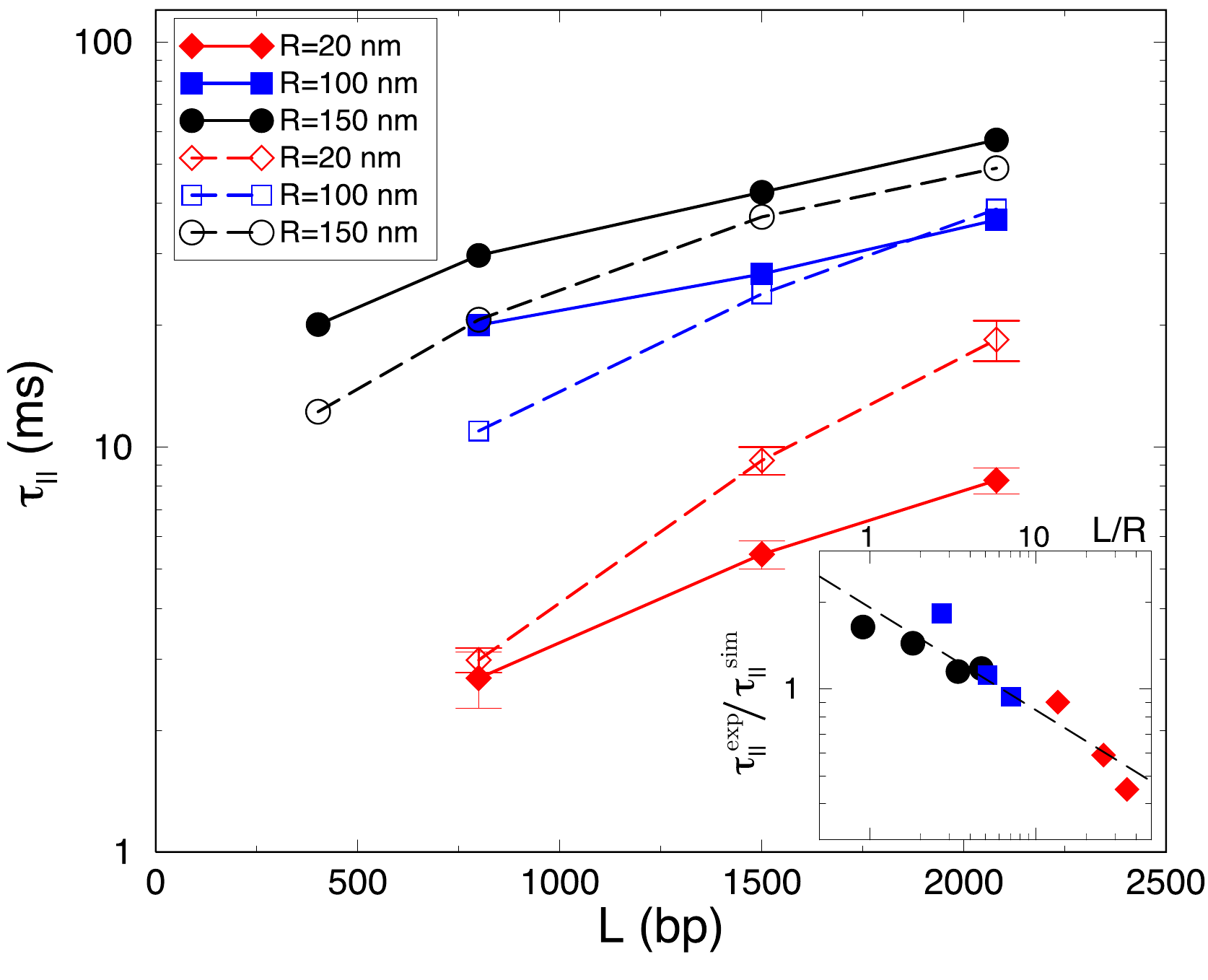}
\caption{\footnotesize Experimental ($\tau_{\parallel}^{\rm exp}$, solid symbols) and numerical ($\tau_{\parallel}^{\rm sim}$, using $z$-dependent diffusion coefficients, open symbols) relaxation times for different DNA lengths, $L$, and particle radius, $R$, in linear-log coordinates. Inset: Ratio $\tau_{\parallel}^{\rm exp}/\tau_{\parallel}^{\rm sim}$ versus $L/R$ with the same symbols as above, in log-log coordinates. The dashed line shows the best linear regression, with slope $-0.355$. Taken from Ref.~\cite{Manghi2010}.}
\label{tau:compare}
\end{center}
\end{figure}

The advantage of these simulations is to give access to dynamical properties.  {From this viewpoint, Brownian Dynamics and kinetic Monter Carlo simulations are equivalent in the small $\delta t$ limit.} As an example, experimental and numerical values of the relaxation times $\tau$ (extracted from $C(t)$ defined in \eq{Ct}) are shown in Figure~\ref{tau:compare}~\cite{Manghi2010}. Experimental and numerical values are found in good agreement, with ratios of experimental to numerical values varying from 0.5 to 2 (shown in the inset).

If one is only interested in equilibrium properties of the conformation of the DNA--particle complex, i.e. the probability distribution $p(|\br_{\parallel}|)$ or its standard deviation $\sigma$ for various apparent DNA contour lengths $L$, a faster numerical method is to compute DNA--particle conformations by exact Monte Carlo sampling. This method has been developed by Segall and collaborators~\cite{Segall2006,Nelson2006} and used in subsequent works~\cite{May2014,Brunet2015}. It consists in generating labeled DNA as a random walk of $N$ steps with a bending energy $E_b$ (defined in Section~\ref{model}) by step. {The angles $\theta_i$ between successive links are randomly chosen through a probability distribution in 
agreement with the Boltzmann weight at equilibrium $\propto \exp[-E_b/(k_BT)]$.} The starting point is {the bead tethered} to the substrate, and at each step, self-intersecting trajectories (resp. trajectories intersecting the substrate) are discarded to take into account intra-chain excluded volume interactions (resp. repulsive interactions with the substrate). Then statistical averages are computed. This numerical method made possible comparisons with approximate analytical expressions, and the finding of {very accurate} interpolation functions for $\sigma(L)$~\cite{Nelson2006}. Moreover, it has been used in Ref.~\cite{Brunet2015} to obtain a graphical reference for $\sigma$ {in function of the dsDNA persistence length $\ell_p$} and therefore to study quantitatively the variation of $\ell_p$ with the salt concentration in the solution (see Section~\ref{salt}).

\subsection{Examples}
\label{ex:sec}

Weak magnitude changes in DNA conformation can be difficult to detect {by TPM}. Coupling multiplexed single-molecule experiments, statistical physics and mesoscopic modeling allows one to detect and investigate narrow changes in the amplitude of movement $\sigma(t)$. Local modifications along the DNA molecule, such as kink, protein binding, loop formation, or more global effects due to a change in the surrounding environment, such as viscosity, $p$H, ionic strength or temperature, will impact the physical properties of the DNA molecule. Changes in the distribution of DNA molecule conformations induce a transition in the apparent contour length, a variable directly accessible through single-molecules techniques. We now review representative examples. 

\subsubsection{Intrinsic curvature angle in TPM}

Local bending of the DNA double helix axis can be induced by either the binding of proteins~\cite{Pouget2006,Zocchi2006,Gietl2013} (Figure~\ref{TPMsketch}, right) or by specific sub-sequences. Specific sub-sequences, such as short A-tracts composed of a succession of adenines {on the same strand}, can locally change the bio-molecule mechanical properties, which is measured through small local bends. Different single-molecule techniques can give access to quantitative measurement of local bending, including AFM, fluorescence spectroscopy, tweezers and TPM~\cite{Tolic2006,Gietl2013}. 

Joint theory and simulation establish the adapted formalism to explore the effect of a local bend in TPM experiments. Using the worm-like chain model to describe a local bending deformation, modeled as a kink of angle $\theta\neq 0$ located at distance $\ell$ from one end, the end-to-end distance $R_{\rm ee}$ is given by a modified version of \eq{KP}~\cite{Brunet2015}, the so-called kinked worm-like chain model:
\be
R_{\rm ee}^2= 2\ell_p^2\left[\left(\frac{L}{\ell_p}-2+e^{-\ell/\ell_p}+e^{-(L-\ell)/\ell_p}\right)+\cos(\theta)\left(1-e^{-\ell/\ell_p}-e^{-(L-\ell)/\ell_p}+e^{-L/\ell_p}\right)\right]
\label{cos}
\ee
From a numerical perspective, simulated TPM is adapted to model locally bent DNA by incorporating the preferred angle $\theta$ between three successive beads into the bead chain. Only the inserted sub-sequence is expected to induce an intrinsic curvature, the remainder of the DNA sequence is assumed to be a random one, without any intrinsic curvature. 
For 575-bp-long DNA molecule, an angle of $\theta = \pi$ would induce a decrease of the apparent DNA contour length of $\approx 30\%$ (absolute reduction of $\approx 35$nm) for a bend located in the middle of the DNA compared to DNA without any curvature. In contrast, the variation is of $\approx 15\%$ (absolute reduction of $\approx 15$nm) for a the same bending site now placed at $1/3$ from the DNA free end. 

HtTPM experiments were performed on a set of designed DNA-molecules containing increasing number of A-tracts, from 0 to 7~\cite{Brunet2015B}. Then, the apparent end-to-end distance of the entire DNA molecule was compared to the predictions of the kinked worm-like chain model to extract the bending angle. The discrepancy between experiments and the analytical expression pointed out the fact that the statistical model did not properly incorporate some biological heterogeneity, notably the non-zero intrinsic curvature spread over the whole molecule. Taking it into account leads to a better extrapolation of data. {This highlights a fact often neglected in theoretical approaches: the zero-temperature limit of real DNA is not a straight, regular helix, but rather a slowly meandering one related to the sequence.}

The analytical models can also consider separately the cases of a bend with a fixed angle and of a local flexible hinge with no spontaneous curvature. Those two distinct causes similarly affect the apparent end-to-end distance measured by TPM and are not easy to disentangle. To account for these two mechanical modifications simultaneously, more precise theoretical developments are needed.

\subsubsection{DNA looping in TPM}
\label{looping1}

DNA looping is a common phenomenon, useful to gene regulation in both prokaryotes and eukaryotes. Probing protein binding and the induced loop formation (\cite{Tardin2017} and references therein) helps to resolve more precisely the geometry of the DNA-protein complexes involved in many biological processes. TPM measurements can be used to unravel the structure of the loop. Moreover, by knowing the length of the loop and the positions of protein-binding sites, the change in apparent end-to-end distance measured by TPM can be used to infer geometrical or conformational alteration of the looped protein-DNA structure. Using statistical mechanics models based upon elastic interactions in small DNA molecules, Biton et al.~\cite{Biton2014} and Johnson et al.~\cite{Johnson2014} probed the interactions of DNA molecules with Lac repressor proteins. They performed TPM measurements to extract the distribution and changes in the apparent length of tethered DNA in function of the operators. Due to the symmetry of the two identical dimeric arms of the Lac repressor, different operators can bind to each arm with different possible orientations, yielding distinct loop types, either parallel or anti-parallel.

Monte Carlo simulations were performed~\cite{Biton2014}, where the DNA molecule was modeled as a necklace of rigid beads separated by rigid cylinders, which takes into account the fluctuations of the Lac repressor-mediated DNA looped segment. Then, the looping probabilities for a specific DNA loop configuration were calculated from the simulations. This information gave access to the {probability} of a loop of a particular length and a given topology. Comparing computational results to TPM experimental results enabled the authors to identify the looped state topology. In addition, simulations were also used to estimate the dissociation constants associated with the binding of the Lac repressor to each of the operators. 
This numerical study~\cite{Biton2014} suggests different looping probabilities for anti-parallel {and parallel loop types formed along a 1632~bp-long DNA molecule, to which a particle of radius 160~nm is attached. The looping probabilities for the anti-parallel loop between binding site centers located at base-pairs 444 and 1044 (resp. 1344) are 4-fold (resp. 6-fold) higher than those for the parallel loop.} It also suggests that the anti-parallel loop is entropically favored.

\subsubsection{Effects of salt on elastic properties}
\label{salt}

The amplitude of movement $\sigma$ not only  depends on the physical properties of the DNA-particle system, but also on the physicochemical properties of the surrounding solution. 

The stiffness properties of nucleic acids molecules depend on the solution ionic strength, {defined as $I=\frac12\sum_i z_i^2c_i$ where $z_i$ and $c_i$ are respectively the valency and the concentration of ion $i$}, that induces screening of the electrostatic repulsion between the negatively charged phosphate groups along the sugar-phosphate backbones. Single-molecule techniques permit to characterize changes of the polyelectrolyte mechanical properties induced by changes in ionic strength as pioneered by Lambert \textit{et al.}~\cite{Lambert2006}. Studies based on high-throughput TPM investigated the dependence of the persistence length on salt concentration for monovalent (Li$^+$, Na$^+$, K$^+$) and divalent (Mg$^{2+}$, Ca$^{2+}$) metallic valent ions~\cite{Brunet2015,Guilbaud2019}. In the first study~\cite{Brunet2015}, after correcting the blurring effect using Eq.~\eqref{corr:time}, the end-to-end distance of the tethered 1201 and 2060~bp-long dsDNA was observed to decrease as a function of the ionic strength $I$ (ranging from 10 mM to 3 M), as expected. A TPM coarse-grained model, taking into account excluded volume interactions, was used to extract $\ell_p$ from measurements by resolving the inverse problem (described in Section~\ref{MC:MD}). It was observed that $\ell_p$ varies from 30 to 55 nm over the large range of ionic conditions, comparable to previous experimental results. 

Two main models were used to fit the data: the Odijk-Skolnick-Fixman model valid at large $I$, relying on a mean field approach valid at low values of the electrostatic potential~\cite{Odijk1977,Skolnick1977}; the Odijk-Skolnick-Fixman-Manning model where the above approach was corrected at low $I$ accounting for the Manning condensation of a few counterions that decreases the effective charge along the DNA~\cite{Manning1981}.
These models could not account quantitatively for the whole experimental data set obtained with Na$^+$ or Mg$^{2+}$.
The more recent Manning model~\cite{Manning2006} with internal electrostatic stretching force due to the repulsion of the charges along the polyelectrolyte well fitted the entire range of $I$ for the Na$^+$ case only. 

The second experimental study~\cite{Guilbaud2019} refined the physical understanding of the phenomenon by exploring the role played by ion size and a larger range of $I$ (from 0.5~mM to 6~M). The experimental protocol was also improved because $p$H was observed to decrease significantly in phosphate buffer when ions were added. A 4-(2- hydroxyethyl)-1-piperazineethanesulfonic acid (HEPES) buffer was used instead and a  slower decrease of the apparent length of the dNA molecule was now observed as comparer to the previous study. Then, the extracted persistence length of the DNA could be quantitatively described by {more sophisticated theoretical approaches, the Netz-Orland~\cite{Netz2003,Brunet2015} for divalent ions and Shen-Trizac~\cite{Trizac2016} for monovalent ones}. These theories, by including non-linear electrostatic effects and the finite DNA radius, can account for the observed behavior of $\ell_p$ over the whole $I$ range. Interestingly, the metallic ion size does not influence the persistence length in contrast to alkyl ammonium monovalent ions at high $I$~\cite{Guilbaud2019}.

A recent work also probed the salt dependence of the torsional stiffness of DNA by multiplexed MTT~\cite{Kriegel2017}. Few different Na$^+$ monovalent concentrations, $20$, $100$ and $500$~mM, and a combination of $100$ mM of Na$^+$ and  $10$ mM of Mg$^{2+}$ were tested. The extension-rotation and torque-rotation curves were collected. The effective torsional stiffness, $C_{\rm eff}$, was determined by fitting the linear torque-rotation regime for each of the ionic strength conditions over various stretching forces. At high stretching forces ($f > 6$~pN), when stretching forces suppress bending along the DNA, the intrinsic torsional stiffness is independent of salt concentration. However, at small stretching forces, $C_{\rm eff}$ increases when the ionic strength increases. Coupling MTT measurement and simulation of the twistable worm-like chain model, permits to examine in more detail the torsional persistence length of the DNA molecule~\cite{Nomidis2017}. The discrepancy between experimental and numerical measurements underlay that bending and twisting are intrinsically coupled in DNA molecule because of the difference between the major and minor groves. {The bending elasticity is not isotropic anymore.} This effect can be implemented in the alternative twistable worm-like chain elastic model proposed by Marko and Siggia~\cite{Marko1994}, where twist-bend coupling is fully taken into account. The systematic deviations of the twist response of dsDNA investigated by magnetic tweezers experiments with the numerical model reported in previous studies could be explained by taking into account this direct coupling between twist and bend deformations. 

\subsubsection{Effects of temperature{~-- DNA denaturation}}
\label{TvsLp}

Due to base-pairing and stacking energies on the order of  {the thermal energy} $k_B T$, DNA flexibility is strongly dependent not only on the ionic strength, but also on the temperature $T$. It affects the cohesive interactions between the DNA bases as well as the contribution of the chain configurational entropy in the free energy~\cite{Manghi2009}. From a biological perspective, various species live in extreme environments and are subjected either to high temperatures or to large temperature fluctuations. This emphasizes the importance of knowing how DNA structure, properties and protein-DNA interactions are affected by temperature.
To this purpose temperature-controlled TPM studies were performed during the last decade~\cite{Driessen2012,Brunet2018}. This technique allowed one to explore the temperature-dependence of the apparent DNA persistence length $\ell_p$. 

In the measured temperature ranges,  from 23 to 52$^{\circ}$C in Ref.~\cite{Driessen2012} and  from 15 to 75$^{\circ}$C in Ref.~\cite{Brunet2018}, a correlation between the increase of $T$ and the decrease of the apparent end-to-end distance of the DNA molecule was revealed. Driessen \textit{et al.}~\cite{Driessen2012} used a numerical procedure, by solving the Langevin equation, \eq{BD}, to model the Brownian motion of the tethered particle. The simulated results in function of the DNA persistence length were fitted with a quadratic function in order to extract the relation between $\ell_p$ and the amplitude of movement $\sigma$. This empirical equation was used to extrapolate the persistence length from the values of $\sigma$ measured in TPM experiments. As expected, $\ell_p$ is slightly dependent on the AT/GC base-pair composition of the DNA. More surprisingly, this study revealed that the intrinsic flexibility of dsDNA strongly and linearly depends on temperature in a range well below the DNA melting temperature, an effect much stronger than expected. 

Brunet \textit{et al.}~\cite{Brunet2018} investigated further the same question by coupling htTPM experiments and Monte Carlo simulations (Section~\ref{MC:MD}). The extracted values of $\ell_p$ showed a slower decrease of the amplitude of movement as compared to the previous study. Considering the changes in buffer viscosity with $T$, the authors put forward that the detector time-averaging blurring effect (Section~\ref{blur}) needed to be cautiously corrected. The observed decrease of the apparent end-to-end distance of the tethered DNA well below the DNA melting temperature is mainly due to this effect. After carefully correcting the TPM measurements from Ref.~\cite{Driessen2012}, the variation of $\ell_p= \kappa(T)/(k_B T)$ with $T$ is sharper, in much better agreement with the expected dependency of the bending modulus $\kappa$ with the temperature at physiological salt conditions. Up to $60^{\circ}C$, the extracted values of $\ell_p$ display a temperature dependency that can be associated with an intact dsDNA molecule, without a significant fraction of denaturated base-pairs. 

Additional work focussed on the temperature dependence of the response of DNA to torsion~\cite{Kriegel2018}. This study combined single-molecule magnetic tweezers measurements with all-atom molecular dynamics and coarse-grained simulations. DNA extension-rotation curves were measured over a temperature range of 24 to 42$^{\circ}$C. Increasing temperature systematically shifted the extension-rotation curves to a negative number of turns. In other words, the point where the DNA molecule is torsionally relaxed changes  linearly with  temperature. Measurements show that the temperature-dependent helical change is not force-dependent for stretching forces $<1$~pN (the overall extension-rotation response of DNA is symmetric around zero turn at low force, see below). Averaging over small forces gives the DNA helical twist constant $\Delta {\rm Tw}(T)= (-11.0\pm 1.2)^{\circ}$/($^{\circ}$C.kbp), in agreement with anterior studies. Using the oxDNA coarse-grained model, describing DNA as two inter-twined strings of rigid nucleotides, a 600~bp-long DNA molecule is simulated including Debye-H\"uckel screened electrostatic {interactions, at $I=150$~mM} to match experimental conditions. Simulated temperatures ranged from 27 to 67$^\circ$C. The mean twist angle for zero torque decreased as when the temperature was raised. Linear fit over temperature yielded the slope $\Delta {\rm Tw}(T)= (-6.4\pm 0.2)^{\circ}$/($^{\circ}$C.kbp), smaller than the experimental value. Additionally, this question was addressed in all-atom simulations of a 33~bp mixed DNA sequence with explicit water molecules and ions at five different temperatures ranging from 7 to 47$^{\circ}C$. The twist angle linearly decreases in the range, DNA-twist changes are equal to $\Delta {\rm Tw}(T)= (-11.1\pm 0.3)^{\circ}$/($^{\circ}$C.kbp), in close agreement with experimental values. The discrepancy in coarse-grained simulations suggests an important role of structural local changes along the DNA molecule, only taken into account in the all-atom simulation. Coupling experimental results with theoretical predictions, this work suggested that the temperature-dependent change in twist is predominantly due to partial and local loss of hydrogen bonds over the DNA backbone that is not correctly considered in coarse-grained models.

\begin{figure}[t!]
\begin{center}
\includegraphics[width=0.48\linewidth]{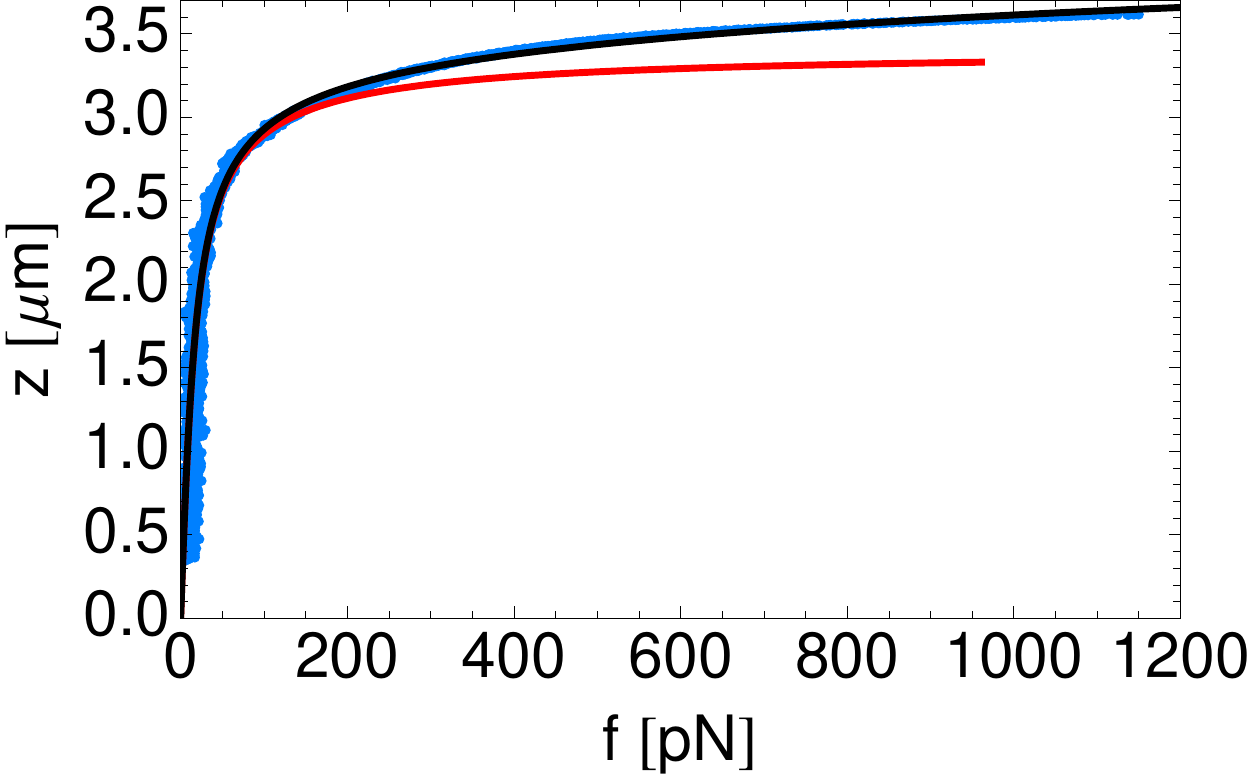}\hfill\includegraphics[width=0.47\linewidth]{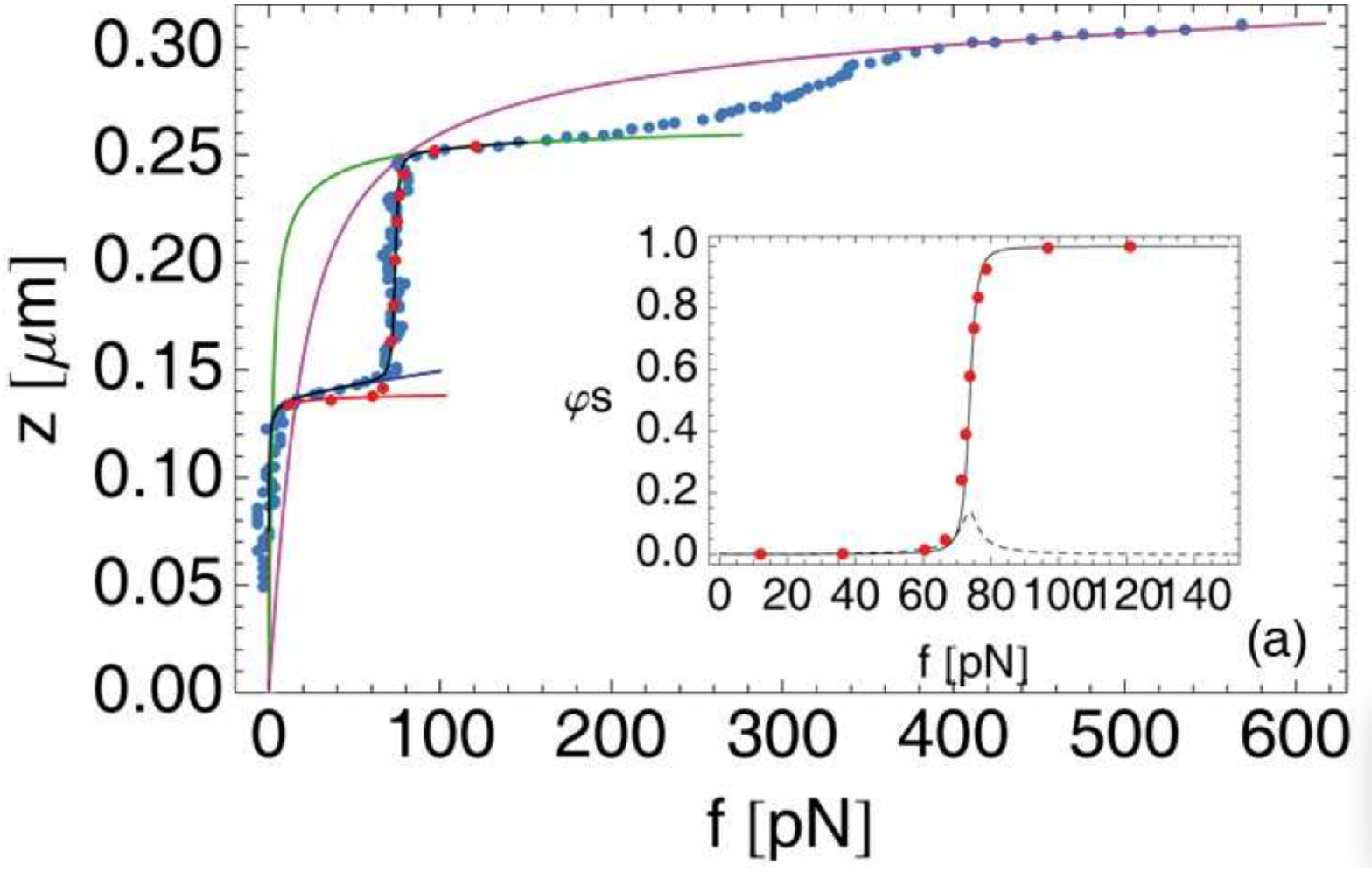}\\
\  {\bf (a)} \hfill {\bf (b)}  \
\end{center}
\caption{\textbf{(a)}~Force-extension curve for a ssDNA. Data (symbols) are taken from H\"ugel \textit{et al.}~\cite{Hugel2005}. The black solid curve corresponds to a fit using the discrete worm like chain interpolation with the non-linear bond elasticity, \eq{ssDNA} (the red one corresponds to discrete version of the Marko Siggia interpolation, \eq{MSF}). The parameters values are: $L=3.40 \;\mu$m, $\tk=1.5$, $a=0.20$~nm. \textbf{(b)}~Force-extension curve for a poly(dG-dC) dsDNA. Data (blue symbols) are taken from Rief \textit{et al.}~\cite{Rief1999}. Solid curves correspond to the discrete worm-like chain interpolation for B-DNA (red), S-DNA (green) and with non-linear extensibility for ssDNA (pink). The black curve corresponds to \eq{extbond}, where linear stretching is included as shown by the blue curve for pure B-DNA. The red symbols correspond to the semi-analytical calculation using transfer matrix. Parameters values are: $L_B = 0.14~\mu$m, $\tk_B = 147$, $\gamma = 1.89$, $\tk_S = \tk_{BS} = 3.8$, $E_B = 1200$~pN. Inset: Fraction of base-pairs  $\varphi_{\rm S}$ in the S state vs. force, and Ising correlation function $1-\langle\sigma_i\sigma_{i+1}\rangle$ (dashed curve). Taken from Ref.~\cite{Manghi2012}.
\label{extensionADN}}
\end{figure}

{DNA structure modifications can be attained not only through temperature changes as discussed above, but also by  applying torque and/or force with tweezers. At low longitudinally applied forces ($f<1$~pN), both strong over- and under-twisting lead to the formation of plectonemic supercoils. At higher forces, plectonemic supercoils are formed under positive torque, but the two DNA single strands separate locally (denaturation bubbles) when a sufficient negative twist is applied~\cite{Vilfan2009}. The consequence of a force applied longitudinally without any external torque is more subtle.} Analytical models are essential tools to determine the nature of the DNA overstretched state observed for $f $ above $\approx 60-70$~pN.  Since its initial discovery in 1996~\cite{Cluzel1996,Smith1996} a debate has arisen as to whether this overstretched state is a new S-DNA form or more simply a denaturated state, i.e. a large denaturation bubble if the two ends are closed or peeled ssDNA if one end is open~\cite{Zhang2013}. New experiments have then been done to probe (i)~the impact of the experimental conditions of attachement on the coverslip and the bead~\cite{Zhang2013}, (ii)~the effect of the NaCl salt concentration~\cite{Zhang2012,Zhang2014}, and (iii)~the influence the base content~\cite{Fu2011,Bosaeus2012}, on the overstretched states. It has been clearly shown (see for instance Refs.~\cite{Manghi2012,Zhang2013}) that the knowledge of the different formulas that fit these three possible states are central in the interpretation of data. Although fitting the transitions itself needs complicated theories such as the one presented in {\ref{extension:app}} leading to the fit shown in Figure~\ref{extensionADN}, simpler formulas such as \eq{MSF} with fewer fitting parameters (the DNA length $L$, its persistence length $\ell_p$) or \eq{ssDNA} for ssDNA stretching allows one to undoubtedly recognize the overstretched state.

This issue is an example where analytical approaches cannot be replaced by numerical simulations since a good numerical model would necessitate both the structural details of the double helix and a dsDNA length $L$ between 0.2 and $4~\mu$m (i.e. 600 to 10000~bp). An attempt has been done using the oxDNA code~\cite{Romano2013} for a 100~bp dsDNA but the S-DNA state has not been observed.

\section{Dynamically detecting two (or more) distinct states}

Detecting dynamical configurational changes of DNA molecules is challenging in many situations of  interest in genetic regulation. As already stressed in Section~\ref{looping1},  protein-induced DNA looping is  a paradigmatic mechanism that has drawn much attention during the last 25 years~\cite{Tardin2017}, because single-molecule techniques have enabled various research groups to shed light not only on looping thermodynamics of different molecular systems but also on their kinetics. As TPM minimizes mechanical constraints on DNA and proteins, it can give access to kinetics at the molecular scale, with high time-resolution. In 1995, lactose repressor-mediated loop formation and disassembly were kinetically monitored for the first time~\cite{Finzi1995}. In this case, the total DNA molecule was $L=1150$~bps long, the polystyrene particle radius was $R=115$~nm, the two DNA sites that are bound when the loop is formed were $\Delta s \simeq 300$~bps away, and the repressor concentration was 1~nM. The looping and disassembly lifetimes were  found to  be very long and both on the order of 100~s. These values were later refined~\cite{Vanzi2006}. Systematic exploration of the effect of the distance $\Delta s$ on looping time was performed in Ref.~\cite{Schickinger2018}. In 2006, the IS911 transpososome assembly was analyzed by following a similar strategy~\cite{Pouget2006}. During the last decade, the bridging activity of site-specific recombinases could also be studied by TPM or TFM by employing a construct where the synapse assembly also reduces the apparent length of the DNA molecule by forming a loop~\cite{Pinkney2012,Diagne2014,May2014,Fournes2016,Fan2018}. Further addition of sodium dodecyl sulfate (SDS) allows one to assess whether strand exchanged occurred or not within the synapse.

Other two-state system kinetics have also been thoroughly studied by TPM. When protein binding provokes a DNA bend, it is detected as an effective shortening~\cite{Pouget2006,Tolic2006,Diagne2014,Fan2018}. Nucleosome assembly in eukaryotes also leads to a shorter apparent length~\cite{Fan2015}. Furthermore, TPM can be used to probe the kinetics of (single) secondary bonds, which can, e.g., transiently form between the particle and the substrate~\cite{Merkus2016}. 

TPM is well adapted to follow dynamics of two-state systems when the dwell-times in both states are on the same order of magnitude, in other words when their free energies are comparable. In this section, we explain how recent theoretical developments likely improve the indirect measurement accuracy of the transition rates between the two (or more) states with TPM.

\subsection{Thresholding and correlation time}
\label{threshold}

\begin{figure}[ht]
\begin{center}
\includegraphics[height=6.1cm]{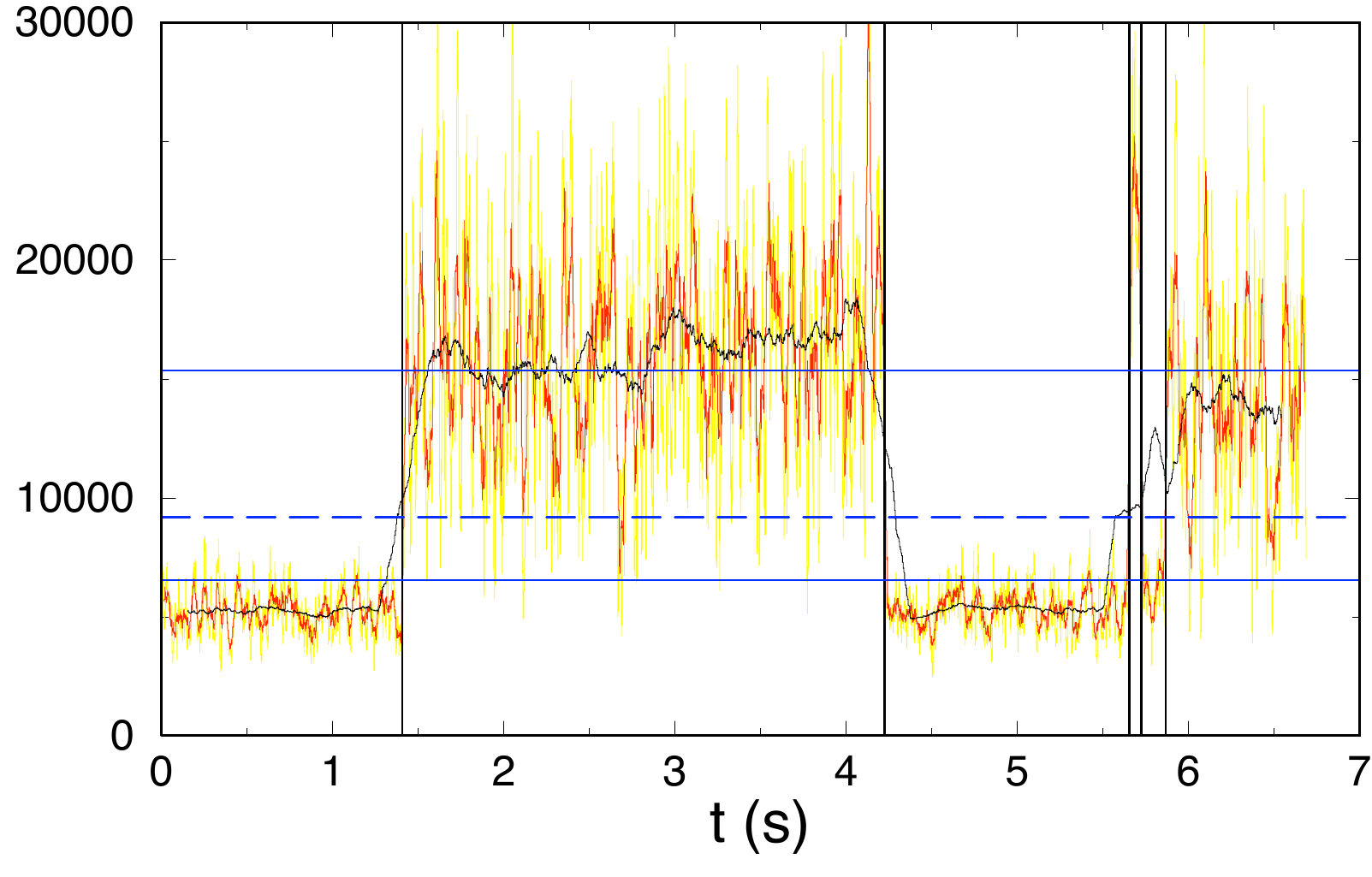} 
\caption{Thresholding: The plots of the variance $\sigma^2$ (in nm$^2$) versus time for a simulated TPM experiment of a DNA molecule of length $L = 798$~bp, a particle of radius $R = 20$~nm and three averaging times $T_{\rm av}= 3$ (yellow), 30 (red) and 300~ms (black). The vertical lines indicate the transition events (looping or unlooping) that are forced in the simulation~\cite{Manghi2010}. The horizontal solid lines show the average values of $\sigma^2$ in the looped (bottom) and unlooped (top) states. The dashed  horizontal line shows the threshold value $\sigma_c$ separating these two states for detection purposes. Detecting close-lying transitions, as in the right part of the figure, is the most critical issue of thresholding together with detection of false transition events (see text). Taken from Ref.~\cite{Manghi2010}.
\label{threshold:fig}}
\end{center}
\end{figure}

We  illustrate the concept of thresholding~\cite{Dixit2005,Laurens2009,Manghi2010} (Figure~\ref{threshold:fig}) in the case of DNA looping between two specific operators and mediated by DNA-binding proteins~\cite{Finzi1995,Vanzi2006,Pouget2006,Johnson2014,Schickinger2018}.  The average dwell-times in the unlooped and looped states are respectively denoted by $\tau_{\rm LF}$ and $\tau_{\rm LB}$. By definition, the transition rates between these two states are $\tau_{\rm LF}^{-1}$ and $\tau_{\rm LB}^{-1}$. 
These quantities depend on the binding energy of the DNA-protein complex, on the protein concentration, and on the DNA elastic properties~\cite{Vanzi2006}. We assume that  the slowly diffusing bead does not significantly alter looping kinetics if it is sufficiently small, as discussed in Ref.~\cite{Manghi2010} {and in Section~\ref{extension}}. 

The most basic idea is to start from the fact that the amplitude of movement $\sigma$ of the tethered particle is smaller in the looped state. Hence the plot of $\sigma$ (or the variance $\sigma^2$)  in function of time will display an alternance of time intervals where $\sigma$ is small and large, as displayed in Figure~\ref{threshold:fig}. On must define a threshold, denoted here by $\sigma_c$, below which the molecule is considered to be looped, and above which it is unlooped. The value of $\sigma_c$ can be set by examining the bimodal distribution of amplitudes of movement~\cite{Vanzi2006,Pouget2006}. One can for example set it at half-amplitude between the two maxima of the bimodal distribution. 
Since $\sigma^2 \equiv \langle \mathbf{r}_{\parallel}^2 \rangle$, the average $\langle \ldots \rangle$ must be estimated on a (sliding) time-window, the duration of which, also called the averaging time, is denoted here by $T_{\rm av}$. From a signal-processing perspective, this averaging scheme corresponds to window-filtering, i.e. convolution of $\mathbf{r}_{\parallel}^2(t)$ with a window function. One might choose to switch to more elaborate exponential or Gaussian filters~\cite{Vanzi2006,Vanzi2007,Colquhoun}, without gaining a significant advantage, however. 

One of the main difficulties comes from the fact that the probability distributions of $\sigma$  for the looped and unlooped states can overlap substantially when using a too short averaging time $T_{\rm av}$~\cite{Beausang2007}, as illustrated in Figure~\ref{threshold:fig}. However, only events occurring at a time-scale larger than $T_{\rm av}$ can be detected. Efficient thresholding thus relies on a compromise between a large value of $T_{\rm av}$ needed to estimate at best the amplitude of movement (and avoid at best false detections), and a short value needed to get the best time resolution (and minimize missed transition events)~\cite{Vanzi2007,Manghi2010}. In particular, the measured values of the rate constants can depend significantly on the window size~\cite{Beausang2007} because the rate is the inverse of the average dwell-time in a state, and measured dwell-times are bounded below by the window size. Consequently, the $T_{\rm av}$ must be chosen consistently with the dwell-times,  typically $T_{\rm av} < \tau_{\rm LF}, \tau_{\rm LB}$, even though some improvements can substantially correct for missed events~\cite{Vanzi2007}.

A lower bound on $T_{\rm av}$ comes from the correlation time $\tau$ introduced in Section~\ref{blur}, as discussed in Ref.~\cite{Manghi2010}. Let us assume for simplicity that the two states have comparable correlation times $\tau$. Their amplitudes of movement are $\sigma_1 < \sigma_2$. We introduce the parameter $\lambda = 1 - (\sigma_1 / \sigma_2)^2$. It was demonstrated~\cite{Manghi2010} that the minimal averaging time  $T_{\rm av}$ needed to resolve them with good accuracy (i.e. with few false detections of transitions) is typically equal to $\tau /\lambda^2$. All in all, optimal $T_{\rm av}$ must satisfy
\begin{equation}
\frac{\tau }{\lambda^2} <  T_{\rm av} < \tau_{\rm LF}, \tau_{\rm LB} .
\label{threshold:limits}
\end{equation}

\subsection{Hidden Markov chains}

Hidden Markov modeling (HMM) combined with a maximum-likelihood approach can be used to determine the numerical values of model parameters such as the transition rates. In contrast to thresholding, no windowing is required, nor the prior selection of a threshold. HMM, initially developed by mathematicians~\cite{Baum1966}, has nowadays plenty of applications in various fields of science. It has been adapted to TPM in 2007 by Beausang \textit{et al.}~\cite{Beausang2007,Beausang2007A}. The underlying idea is that the system under consideration can be modeled by a Markov chain~\cite{Grimmett}, the states of which are not directly observed in the experiment, i.e., they are ``hidden''.

Here we again illustrate these ideas in the case of DNA looping, even though it can be generalized, e.g., to protein-binding. The hidden state, denoted by $q(t)$, can be ``looped'' or ``unlooped'' DNA. We again look for the average dwell-times $\tau_{\rm LF}$ and $\tau_{\rm LB}$. The most basic idea~\cite{Beausang2007} would be to consider that this two-state system is governed by a two-state Markov chain with a $2 \times 2$ transition matrix~\cite{Grimmett}. However, this naive approach fails because it ignores the fact that transition rates between both states depend on the polymer conformation: for example, looping is forbidden when the polymer is too stretched. As a consequence, the Markov chain must also take into account the chain configuration, for example through the 2D position of the particle, $\mathbf{r}_{\parallel}$, itself governed by an over-damped Langevin equation in a harmonic potential.  The system state as it appears in the Markov chain is now $(q(t),\mathbf{r}_{\parallel}(t))$. 

Once an experimental time series $(q(t),\mathbf{r}_{\parallel}(t))_t$ has been recorder, the idea is then to calculate the likelihood that it is observed for a given pair of dwell-times $(\tau_{\rm LF},\tau_{\rm LB})$. This can be done with the standard tools of probability theory. Then the dwell-time values maximizing this likelihood are considered to be the most probable ones. This procedure has been tested on numerically generated trajectories for which the dwell times were exactly known. It was able to correctly recover these values, up to statistical error bars.

However, in spite of its conceptual simplicity, the practical implementation of the method relies on some strong approximations about the looping process, as stated by the authors themselves~\cite{Beausang2007}. For example, looping is allowed if and only if the particle excursion  $\| \mathbf{r}_{\parallel}(t) \|$ is smaller than a threshold $\rho_{\rm max}$. This is modeled by  a crude $\Theta$ step-function in the Markov chain. In addition, a simplification is ``to  ignore the unobserved height variable $z$ [above the glass substrate plane], in effect treating the bead motion as diffusion in two dimensions.''  This is again an issue when deciding whether looping is possible or not for a given value of $\| \mathbf{r}_{\parallel}(t) \|$ because $z$ can be large even though $\| \mathbf{r}_{\parallel}(t) \|$ is small. ``A better analysis might treat $z$ as another hidden (unobserved) variable.''  Finally, as in the thresholding approach, the polymer is considered to be in quasi-equilibrium in both the looped and unlooped state, from which transition probabilities are inferred. This assumption is only valid if the free-energy wells around each state are sufficiently deep. 

A refinement of the HMM method relies on variational Bayesian inference, first used in the study of Lac repressor-mediated looping~\cite{Johnson2014} that we have already mentioned in Section~\ref{looping1}. Bayesian inference is able to determine not only the most likely model parameters but also the most likely number of hidden states, that was fixed \textit{a priori} in the above HMM method, by observing the number of peaks in the probability distribution of $\sigma$. The improved ability of this method to resolve close-lying transitions (see Figure~\ref{threshold:fig}) has also been tested in this work.  For instance, two states separated by an amplitude of movement $\sigma$ of 40 nm could be resolved by their technique at a mean lifetime of about 0.5~s. In contrast, lifetimes of more than 4~s were necessary for states to be resolvable by simple thresholding. The reason for this difference is that Bayesian inference does not require time-filtering as evoked above. 

When the free energies of the states are significantly different, one state is favoured with respect to the other. For example, let us assume that once the loop is formed, it is very stable because breaking it would require a free energy $\Delta F \gg k_{\rm} T$. Then loop opening events become rare and cannot be observed in practice. It becomes interesting to apply a force $f$ {(with the help of tweezers)} in the pN range on the tethered particle because the associated potential energy difference between both states can compensate $\Delta F$ and make them roughly equiprobable, thus allowing one to observe more frequent opening events. Indeed, the dwell-times are known to depend exponentially on the applied force~\cite{Evans1997}. 

Beyond looping, this approach has been successfully used in combination with HMM analysis to study single-nucleosome unwrapping~\cite{Kruithof2009}. Since DNA is wrapped in about two turns around a histone octamer, three distinct conformations can be observed: fully wrapped, about one turn unwrapped and fully unwrapped. Applying a 2.5~pN force allows one to observe transitions between the two first states, and at a stretching force of 6 pN, the second turn unwraps. Detailed information about nucleosome unwrapping such as bending angles of nucleosomal DNA or dwell-times could be extracted from these experiments. Zero-force dwell-times can then be extrapolated from these measurements. This method can in principle be used to detect any transient DNA-protein complex formation and dwell-times. Similar approaches have been used to study DNA hairpins opening/closure~\cite{Manghi2016} and DNA G-quadruplex unfolding/refolding~\cite{You2018}.

\section{Conclusion}

This Review has illustrated in many situations the powerful capabilities of {TPM} experiments coupled to theoretical and/or numerical modeling to give access to quantities of interest in both biological and biophysical contexts. We have identified two levels of difficulties that must be overcome in order to infer the physical parameters of interest with a good accuracy. 

First, raw data must be carefully processed thanks to well-established protocols in order to deal with both statistical and systematic sources of errors. The existence of outliers seems inherent to single molecule experiments because it is both difficult to prepare samples with 100~\% of identical molecules and to graft all molecules in ideal conditions limiting unwanted non-specific interactions. This is in part solved thanks to nano-lithography techniques, but not entirely. After having discarded outliers, the most critical systematic effect to deal with is the blurring effect inherent to the finiteness of the detector exposure. We have proposed an efficient way to solve this issue through  simple inversion formulae, as summarized in Eqs.~\eqref{corr:time}  and  \eqref{ell:ellm0}.

Once the data have been corrected from these sources of error, solving the inverse problem enables one to infer the DNA state or the physical parameters of interest. This concerns  not only quantities measured at thermodynamic equilibrium (e.g., elastic parameters $C$ and $\ell_p$ or intrinsic bending angles), but also out-of-equilibrium properties of great biological interest such as transition rates (e.g. binding/unbinding or looping/unlooping rates). In the latter case, we have shown that hidden Markov chain approaches are promising even though  they are more complex to implement than simple thresholding. In all cases, the underlying quantitative model must rely on solid physical grounds, appealing to polymer and elasticity theory, and out-of-equilibrium statistical mechanics. We have listed several successes of such approaches in this Review. 

However, to our point of view, few issues remain to be solved in the future in order to provide a fully operational tool to biophysicists and biologists. We have explained that intrinsic curvature modifies the amplitude of movement in a way that can be quantified with good accuracy. However, weak intrinsic curvature is spread all over the molecule, and it is not only localized at specific high-curvature loci. This ``quenched'' disorder likely modifies the amplitude of movement in a systematic but ill-controlled manner. This phenomenon should be quantified, for example by using more sophisticated mesoscopic models fully taking account such subtle effects~\cite{Zuiddam2017}. 

When a region of DNA is modified or affected in any way, for example through binding of a protein, it can bear not only a different spontaneous curvature that one wants to quantify, but also a different bending modulus. In particular, DNA flexibility is a sequence-dependent quantity~\cite{Geggier2010}. Disentangling sequence-dependent spontaneous curvature and sequence-dependent elastic properties in single-molecule experiments is another issue that will require extensive modeling work in the future. 

As far as dynamical properties are concerned, we have just stressed that methods relying on hidden Markov chains are quite promising. However, to our knowledge, no systematic quantification of their capabilities, in the spirit of Eq.~\eqref{threshold:limits}, has been performed so far. Probability theory together with numerical modeling should be able to give definitive and robust conclusions on the strengths and limits of these approaches. Filling this gap seems important to us in order to eventually provide an easy-to-use tool to experimentalists.

\section*{Acknowledgments}

We warmly thank our colleagues Catherine Tardin and Laurence Salom\'e for fruitful discussions and sound advices during the writing of this Review.
We are also tributary to the Universit\'e Toulouse III-Paul Sabatier and the Centre National de la Recherche Scientifique (CNRS).

\appendix

\section{Analytical modeling of stretching experiments}
\label{extension:app}

The classical model for semi-flexible polymer stretching has been developed by Marko and Siggia~\cite{Marko1995} using the continuous worm-like chain model by developing a formula that interpolates between the exact results in the limits of low ($f\ll k_BT/\ell_p\simeq 0.08$~pN) and strong forces ($f\gg k_BT/\ell_p$):
\be
f=\frac{k_BT}{\ell_p}\left[\frac{z}L-\frac14+\frac1{4(1-z/L)^2}\right].
\label{MSF}
\ee
This formula has been successfully used to fit the force-extension curves for $f\lesssim65$~pN for a $\lambda$-phage dsDNA (see the review~\cite{Bustamante2000} and references therein).
One technical difficulty in fitting these curves is to set the origin. This might be correlated to the fact that experimentally the tethers are not stretched exactly in the  direction perpendicular to the coverslip~\cite{Ray2007}.

For larger forces, the DNA internal structure starts to come into play (see below). However this type of approach remains valid for very high stretching of ssDNA~\cite{Rief1999} provided that both the discrete nature of the chain is taken into account~\cite{Kierfeld2004} and non-linear stretching terms are included~\cite{Hugel2005}. Fits of force-extension curves of ssDNA up to 1200~pN, as shown in Figure~\ref{extensionADN}a, have been nicely fitted using the modified formula that includes both terms~\cite{Manghi2012}:
\be
f=\frac{k_B T}{a}\left[ \frac{z}{L} (1+U_{\rm nl}(f)) \left(3\frac{1-u(\tk)}{1+u(\tk)}-\frac1{\sqrt{1+4\tk^2}}\right)+\left(\frac1{[1-z(1+U_{\rm nl}(f))/L]^2}+4\tk^2\right)^{1/2}-\sqrt{1+4\tk^2}\right]
\label{ssDNA}
\ee
where $\tk=\kappa/(k_BT)$ is the bending modulus in units of $k_BT\simeq4\times 10^{-21}$~J (at room temperature), $u(\tk)=\coth(\tk)-1/\tk$, and $U_{\rm nl}(f) = 1.172777\; f  - 3.731836\;f^2 +  4.118249\; f^3$ (where $f$ is in units of 10~nN)~\cite{Hugel2005}. The fit leads to the bending modulus value $\kappa(\mathrm{ssDNA})=1.5\ k_BT$ and an effective monomer length $a=0.20$~nm (see Figure~\ref{extensionADN}a). Unexpectedly this value of $a$ is much smaller than the distance between two consecutive bases in ssDNA $a_{ss}\simeq0.7$~nm.
This result suggests that the number of degrees of freedom increases by a factor 3.5 for strong forces~\cite{Manghi2012}, a result that has already been observed for peptides~\cite{Hanke2010}.

When strong forces are applied to dsDNA, experiments show a sharp, few picoNewtons wide, cooperative overstretching ÒtransitionÓ at a given ÒcriticalÓ force of around 60--80~pN , accompanied by a sudden 70\% increase of the contour length~\cite{Cluzel1996,Smith1996}. It corresponds to a transition from the B-form to a new form of unstacked DNA remaining in a duplex form (S-DNA form for Stretched). A second transition is observed at stronger forces~\cite{Rief1999} consistent with a peeling of one strand from the other strand. The critical force and therefore the selection between these two transitions depends on the DNA sequence and the salt concentration~\cite{Fu2010,Fu2011,Zhang2012}.
Using a Ising-Heisenberg coupled model~\cite{Storm2003,Palmeri2007,Palmeri2008}, an analytical formula has been derived~\cite{Manghi2012} which allows us to fit the first transition:
\bea
\frac{z}{a_B N} &=& \left(1+\frac{F}{\tilde E_B}-\frac1{2\alpha_B}\right)\varphi_B+\gamma\left(1-\frac1{2\alpha_S}\right)\varphi_S \nonumber \\
&&+\frac{\langle\sigma_i\sigma_{i+1}\rangle-1}4\left(\frac1{2\alpha_B}\frac{\tk_B-\tk_{BS}}{\tk_{BS}+F/2+\alpha_B}+\frac\gamma{2\alpha_S}\frac{\tk_S-\tk_{BS}}{\tk_{BS}+\gamma F/2+\alpha_S}\right)
\label{extbond}
\eea
where B and S refer to the DNA state, $F=a_Bf/(k_BT)$, $\tk_{BS}$ is the bending modulus at the BS domain wall, $\gamma=a_S/a_B$, $\tilde E_B=a_BE_B/(k_BT)$ where $E_B$ is the stretching modulus, $\alpha_B=(\tk_BF+F^2/4)^{1/2}$ and $\alpha_S=[\tk_S\gamma F+(\gamma F)^2/4]^{1/2}$. The fraction of base-pairs in the B (respectively S) state are $\varphi_{B}$ (respectively $\varphi_S=1-\varphi_B$). Finally $\langle\sigma_i\sigma_{i+1}\rangle$ is the two-point correlation function of the effective Ising model which is non-zero only close to the transition~\cite{Manghi2012}. An example of a fit of the force-extension curve for a poly(dG-dC) DNA is shown in Figure~\ref{extensionADN}b. The inset shows the variation of the fraction of base pairs in the S state and the Ising correlation function as a function of the applied force.
\eq{extbond} has also been used to fit the S-DNA to ssDNA transition for poly(dG-dC) where the linear stretching term is replaced by the non-linear stretching one for ssDNA~\cite{Manghi2012}.


\end{document}